\documentclass[12pt,thmsa]{article}
\usepackage{sw20aip}

\input{tcilatex}
\input tcilatex
\begin{document}

\author{A. Tartaglia \\
Dip. Fisica, Politecnico, Torino, Italy\\
e-mail: tartaglia@polito.it}
\title{Detection of the gravitomagnetic clock effect}
\maketitle

\begin{abstract}
The essence of the gravitomagnetic clock effect is properly defined showing
that its origin is in the topology of world lines with closed space
projections. It is shown that, in weak field approximation and for a
spherically symmetric central body, the loss of synchrony between two clocks
counter-rotating along a circular geodesic is proportional to the angular
momentum of the source of the gravitational field. Numerical estimates are
presented for objects within the solar system. The less unfavorable
situation is found around Jupiter.

\end{abstract}

\section{Introduction}

Recently Bonnor and Steadman\cite{bonnor} (BS) published a paper on the so
called gravitomagnetic clock effect. The subject is important because it
could hint at an opportunity to verify a general relativistic effect caused
by the angular momentum of a source of gravitational field. The present work
is intended to clarify the nature and the origin of the effect and contains
also some new calculations of the synchrony defect between identical clocks
moving in opposite senses around a spinning massive body; the order of
magnitude of this defect is within the range of the observations which can
be made in the solar system.

The gravitomagnetic clock effect BS speak of is the difference in time shown
after one revolution by two identical clocks co-rotating and
counter-rotating on a circular orbit around a spinning mass\cite{cohen}. The
above authors show that the amount of the time difference depends on the
reference frame with respect to which the phenomenon is described and
discuss two special cases: a rotating massive infinitely long dust cylinder
and a Kerr black hole.

In fact however, with a slight difference in perspective, one could better
speak of a gravitomagnetic effect directly comparing the proper times read
on two oppositely rotating clocks when their trajectories intersect each
other at a point. The difference in the readings will so be an absolute
quantity, whose value is a measure of the angular momentum of the source.
Other ''clock effects'' have been considered in the literature depending on
the nature of the clock and related for instance to the coupling of the spin
with the angular momentum of the gravity source\cite{ahlu}: what we are
treating here is the purely geometrical effect, perfectly insensitive to the
nature of the clock.

We shall first remind and show that the special relativistic root of the
phenomenon is in the behaviour of clocks in rotary motion (sect. II), then
in sect. III it will be shown how the same situation is described in a
gravitational non rotating field. Sect. IV deals with the drag effect due to
the presence of a rotating source. In sect. V the van Stockum and Kerr
metrics are considered. Sect. VI contains some estimates valid for the solar
system environment and finally, in sect. VII, we shall draw some conclusions.

\section{Effect in the absence of a gravitational field}

The standard situation is the one where the rotation is considered from a
view point far away from the gravitational source, where the space time is
flat, and supposing that the observer is at rest with respect to the
rotation axis. The, so to speak, zero level case is the one where the two
clocks perform their rotational motion in the absence of any gravitational
field; of course no closed space geodesics exist in this case, a closed
circuit then requires some constraint. When the two clocks move with the
same speed in opposite directions along a circular path the situation is
trivial: whenever they cross each other at a point they turn out to be
synchronous. The situation is different when the angular speeds are
different, no matter whether they are in the same sense or in opposite
senses. This is indeed a peculiarity of special relativity and a special
case of the so called Sagnac effect. The Sagnac effect is usually expressed
in terms of the propagation of light emitted by a rotating source\cite
{sagnac}, but its essence can be reduced precisely to the lack of synchrony
of clocks moving along closed space paths\cite{anandan} (for a comprehensive
review on the Sagnac effect see\cite{stedman}). The phenomenon is a purely
geometrical property of Minkowski space time. In fact the four dimensional
trajectory of a steadily rotating object is a timelike helix\cite{rt}; the
proper times to compare, shown by two identical clocks rotating with
different angular speeds along the same circumference, are the lengths of
the helixes arcs intercepted between two successive intersection events:
these lengths are necessarily different when the speeds are different.

Let us call $R$ the radius of the circular path, $\phi $the rotation angle, $%
\omega $ the angular speed with respect to an inertial static observer and $t
$ the time of the inertial observer. The world line of a rotating clock may
then be written 
\begin{equation}
\left\{ 
\begin{array}{c}
t=\frac{\phi }{\omega } \\ 
r=R
\end{array}
\right.   \label{coordi}
\end{equation}

Consider now two identical clocks rotating respectively with angular speeds $%
\omega _{1}$and $\omega _{2}$at the same $R$, and synchronous in $t=\phi =0$%
; the next encounter between them will happen when 
\begin{equation}
\frac{\phi }{\omega _{1}}=\frac{\phi \pm 2\pi }{\omega _{2}}  \label{eliche}
\end{equation}

where $\phi $is the angular coordinate of the first clock; the $+$ sign
holds when $\omega _{2}>\omega _{1},$ the $-$ one when $\omega _{2}<\omega
_{1}$.

Solving (\ref{eliche}) for $\phi $(or, to say better, $\phi _{1}$) gives the
angular span between two successive intersections of the helixes: 
\begin{equation}
\phi _{1}=2\pi \frac{\omega _{1}}{\left| \omega _{2}-\omega _{1}\right| }
\label{span}
\end{equation}

For the other clock it would be: 
\begin{equation}
\phi _{2}=2\pi \frac{\omega _{2}}{\left| \omega _{2}-\omega _{1}\right| }
\label{spannuno}
\end{equation}

The length of the helix arc spanned by the angle (\ref{span}) or (\ref
{spannuno}), divided by $c$, is the proper time $\tau $ shown by the
corresponding clock: 
\begin{eqnarray*}
\tau _{1} &=&\int_{helix}\sqrt{dt^{2}-\frac{R^{2}d\phi ^{2}}{c^{2}}}=\sqrt{1-%
\frac{R^{2}\omega _{1}^{2}}{c^{2}}}\frac{2\pi }{\left| \omega _{2}-\omega
_{1}\right| } \\
\tau _{2} &=&\sqrt{1-\frac{R^{2}\omega _{2}^{2}}{c^{2}}}\frac{2\pi }{\left|
\omega _{2}-\omega _{1}\right| }
\end{eqnarray*}

The lack of synchrony at the first intersection event is: 
\begin{equation}
\Delta \tau =\left| \tau _{2}-\tau _{1}\right| =\left| \frac{2\pi }{\omega
_{2}-\omega _{1}}\left( \sqrt{1-\frac{R^{2}\omega _{1}^{2}}{c^{2}}}-\sqrt{1-%
\frac{R^{2}\omega _{2}^{2}}{c^{2}}}\right) \right|  \label{lack}
\end{equation}

It must be remarked that $\Delta \tau $ is not, strictly speaking,
registered after one revolution of any of the clocks: the two objects meet
again after more or less than one revolution, according to which of them is
being considered. When $\omega _{1}>0$ and $\omega _{2}<0$ the intersection
corresponding to (\ref{lack}) actually happens a bit after or before a half
revolution.

The quantity (\ref{lack}) is of course zero when $\omega _{2}=-\omega _{1}$.
Suppose one of the clocks (be it the number $1$) is stationary in the
inertial reference frame; now its revolution time is infinite, but this does
not mean that $\Delta \tau $ too is infinite or is undefined. The
intersections between the world lines of the two clocks are still perfectly
recognizable; simply it is $\omega _{1}=0$ and $\Delta \tau =\frac{2\pi }{%
\omega _{2}}\sqrt{1-\frac{R^{2}\omega _{2}^{2}}{c^{2}}}$

Formula (\ref{lack}) expresses an absolute, i.e. invariant, quantity. This
means that the result is the same no matter what the reference frame is. We
can in particular consider the case of an observer steadily rotating along
the same circumference at an (inertial) angular speed $\omega _{o}$. The
synchrony defect between the two moving clocks is still (\ref{lack}) but now
it should be expressed in terms of the parameters relevant for the new non
inertial observer. The revolution time of a clock as measured by him is
indeed: 
\begin{equation}
T=\sqrt{1-\frac{R^{2}\omega _{o}^{2}}{c^{2}}}\frac{2\pi }{\left| \omega
-\omega _{o}\right| }  \label{periodo}
\end{equation}

From (\ref{periodo}) it is possible to find the relation between the angular
speeds as seen by the moving observer and the ones measured in the inertial
frame. One obtains 
\begin{equation}
\omega ^{\prime }=\frac{2\pi }{T}=\frac{\omega -\omega _{o}}{\sqrt{1-\frac{%
R^{2}\omega _{o}^{2}}{c^{2}}}}  \label{omega}
\end{equation}

Inverting (\ref{omega}) to find $\omega $, then substituting in (\ref{lack})
we express the synchrony defect in terms of the parameters of the moving
observer: 
\begin{equation}
\Delta \tau =\frac{2\pi }{\left( \omega _{2}^{\prime }-\omega _{1}^{\prime
}\right) }\left( \sqrt{1-\frac{R^{2}}{c^{2}}\left( \omega _{1}^{\prime }+%
\frac{\omega _{o}}{\sqrt{1-\frac{R^{2}\omega _{o}^{2}}{c^{2}}}}\right) ^{2}}-%
\sqrt{1-\frac{R^{2}}{c^{2}}\left( \omega _{2}^{\prime }+\frac{\omega _{o}}{%
\sqrt{1-\frac{R^{2}\omega _{o}^{2}}{c^{2}}}}\right) ^{2}}\right)
\label{lack1}
\end{equation}

The invariant character of the synchrony defect means also that if it is non
zero in one reference frame it will be so in any other too; in other words
no $\omega _{o}$ value is able to bring $\Delta \tau $ to $0$ when $\omega
_{1}\neq -\omega _{2}$ (this inequality corresponds to $\omega _{1}^{\prime
}\neq -\omega _{2}^{\prime }-\frac{2\omega _{o}}{\sqrt{1-\frac{R^{2}\omega
_{o}^{2}}{c^{2}}}}$).

\section{Effect of the gravitational field}

The special relativistic analysis in the previous section is the basis of
the evaluation of the effect when a gravitational field is present. In fact
the structure of the phenomenon is not different; what is different is the
metric of space time. A point to be stressed is that the relevant events are
those in which the two clocks coincide in space; using these events frees
from the need of a third party deciding when a complete revolution has been
performed. Also the fact that the trajectory is a geodesic turns out to be
irrelevant, being reduced to a special case of the general situation. Of
course the presence of a gravitational field renders closed space geodesics
available, which was not the case in Minkowski space time.

Let us consider first a static spherically symmetric metric. The world line
of an object steadily moving along a circular trajectory in an equatorial
plane of the source is still a helix. The typical line element in flat space
time polar coordinates is: $ds^{2}=c^{2}e^{2\lambda }dt^{2}-e^{2\mu
}dr^{2}-e^{2\nu }r^{2}\left( d\theta ^{2}+\sin ^{2}\theta d\phi ^{2}\right) $%
, where $\lambda $, $\mu $, $\nu $ are functions of $r$ only. The
corresponding equation of the helix is 
\begin{eqnarray*}
\phi &=&\omega t \\
r &=&\text{constant}
\end{eqnarray*}

and the helix arc, in terms of proper time, spanned by the angle $d\phi $ is 
\[
d\tau =\sqrt{e^{2\lambda }dt^{2}-e^{2\nu }\frac{r^{2}}{c^{2}}d\phi ^{2}}=%
\sqrt{\frac{e^{2\lambda }}{\omega ^{2}}-e^{2\nu }\frac{r^{2}}{c^{2}}}d\phi 
\]

On these bases the synchrony defect of two identical clocks rotating with
speeds $\omega _{1}$ and $\omega _{2}$ at the same radius $r$ is 
\begin{equation}
\Delta \tau =\frac{2\pi e^{\lambda }}{\omega _{2}-\omega _{1}}\left( \sqrt{%
1-e^{2\left( \nu -\lambda \right) }\frac{\omega _{1}^{2}r^{2}}{c^{2}}}-\sqrt{%
1-e^{2\left( \nu -\lambda \right) }\frac{\omega _{2}^{2}r^{2}}{c^{2}}}\right)
\label{sferico}
\end{equation}

Again (and trivially because of the symmetry) it can be $\Delta \tau =0$
only when $\omega _{2}=-\omega _{1}$.

Formula (\ref{sferico}) is expressed in terms of parameters of an inertial
static observer; it is however an invariant quantity. If we want the result
from the viewpoint of a steadily rotating observer on the same circumference
as the clocks, we can proceed as in the flat space time case. The revolution
time of a clock as seen by the rotating observer is 
\[
T=\frac{2\pi e^{\lambda }}{\left| \omega -\omega _{o}\right| }\sqrt{%
1-e^{2\left( \nu -\lambda \right) }\frac{\omega _{o}^{2}r^{2}}{c^{2}}} 
\]

and the corresponding angular speed is 
\[
\omega ^{\prime }=\frac{\left( \omega -\omega _{o}\right) e^{-\lambda }}{%
\sqrt{1-e^{2\left( \nu -\lambda \right) }\frac{\omega _{o}^{2}r^{2}}{c^{2}}}}
\]

Solving for $\omega $ and substituting into (\ref{sferico}), then posing $%
e^{\nu -\lambda }r=R$ leads to

\begin{eqnarray*}
\Delta \tau &=&\frac{2\pi }{\left( \omega _{2}^{\prime }-\omega _{1}^{\prime
}\right) }(\sqrt{\frac{1}{\left( 1-\frac{\omega _{o}^{2}R^{2}}{c^{2}}\right) 
}-\frac{R^{2}}{c^{2}}\left( \allowbreak \omega _{1}^{\prime }e^{\lambda }+%
\frac{\omega _{o}}{\sqrt{\left( 1-\frac{\omega _{o}^{2}R^{2}}{c^{2}}\right) }%
}\right) ^{2}}+ \\
&&-\sqrt{\frac{1}{\left( 1-\frac{\omega _{o}^{2}R^{2}}{c^{2}}\right) }-\frac{%
R^{2}}{c^{2}}\left( \allowbreak \omega _{2}^{\prime }e^{\lambda }+\frac{%
\omega _{o}}{\sqrt{\left( 1-\frac{\omega _{o}^{2}R^{2}}{c^{2}}\right) }}%
\right) ^{2}})
\end{eqnarray*}

The presence of the term $e^{\lambda }$, now to be considered as a function
of $R$, manifests the presence of the field.

If the clocks are orbiting along a circular space geodesic the angular
speeds are necessarily the same but in opposite directions, consequently it
is trivially $\Delta \tau =0$.

\section{Angular momentum effect}

A more interesting situation is found when the source of the gravitational
field possesses an angular momentum. In this case the metric is in general
axisymmetric; assuming it is also stationary and has a full cylindrical
space symmetry we can write, using appropriate coordinates: 
\begin{equation}
ds^{2}=fc^{2}dt^{2}-2Kcdtd\phi -ld\phi ^{2}-H\left( dr^{2}+dz^{2}\right)
\label{cilindro}
\end{equation}

where $f$, $K$, $l$, $H$ are all functions of $r$ only.

We limit our analysis to the plane $z=0$. A constant speed circular space
trajectory in that plane will correspond again to a helical world line in
space time ($\phi $ is proportional to $t$, $\omega =d\phi /dt$). The helix
arc spanned by an angle $\phi $ and expressed as the proper time of a clock
is now: 
\begin{equation}
\tau =\sqrt{\frac{f}{\omega ^{2}}-2\frac{K}{c\omega }-\frac{l}{c^{2}}}\phi
\label{arco}
\end{equation}

Formulae (\ref{eliche}), (\ref{span}) and (\ref{spannuno}) continue to hold,
consequently it is: 
\begin{equation}
\Delta \tau =\left| \frac{2\pi }{\omega _{2}-\omega _{1}}\left( \sqrt{f-2%
\frac{K}{c}\omega _{1}-\frac{l}{c^{2}}\omega _{1}^{2}}-\sqrt{f-2\frac{K}{c}%
\omega _{2}-\frac{l}{c^{2}}\omega _{2}^{2}}\right) \right|
\label{cilindrico}
\end{equation}

This result differs from the one obtained by Bonnor and Steadman in that
theirs considers revolution times as seen by an inertial static observer,
whereas here $\Delta \tau $ is the lack of synchrony obtained directly
comparing two clocks when they coincide in space time.

Now the effect is zero, excluding the trivial solution $\omega _{2}=\omega
_{1}$, when 
\[
2\frac{K}{c}\omega _{1}+\frac{l}{c^{2}}\omega _{1}^{2}=2\frac{K}{c}\omega
_{2}+\frac{l}{c^{2}}\omega _{2}^{2} 
\]

i.e. for 
\begin{equation}
\omega _{2}=-\omega _{1}-2\frac{K}{l}c  \label{rotante}
\end{equation}

When the angular momentum of the source (which is in general proportional to 
$K$) is zero the solution comes back to the known $\omega _{2}=-\omega _{1}$.

Again we can express (\ref{cilindrico}) in terms of the variables seen by a
non inertial observer rotating at an angular speed $\omega _{o}$ along the
same circumference as the clocks. The revolution time measured by the
observer is now 
\[
T=\frac{2\pi }{\omega -\omega _{o}}\sqrt{f-2\frac{K}{c}\omega _{o}-\frac{l}{%
c^{2}}\omega _{o}^{2}} 
\]

The angular speed he sees is then 
\[
\omega ^{\prime }=\frac{\omega -\omega _{o}}{\sqrt{f-2\frac{K}{c}\omega _{o}-%
\frac{l}{c^{2}}\omega _{o}^{2}}} 
\]

Solving for $\omega $, then introducing the result into (\ref{cilindrico})
one ends up with a rather messy expression. The operation is simpler in the
translation of (\ref{rotante}), which becomes 
\[
\omega _{2}^{\prime }=-\omega _{1}^{\prime }-2\frac{\omega _{o}+\frac{K}{l}c%
}{\sqrt{f-2\frac{K}{c}\omega _{o}-\frac{l}{c^{2}}\omega _{o}^{2}}} 
\]

The results found up to now can be specialized to the case of geodesic
circular trajectories with the same radius $r$. For such geodesics the
condition must be satisfied\cite{bonnor}:

\[
c^{2}\frac{df}{dr}-2c\frac{dK}{dr}\omega -\frac{dl}{dr}\omega ^{2}=0 
\]

Two solutions exist for $\omega $, provided $\left( \frac{dK}{dr}\right)
^{2}+\frac{dl}{dr}\frac{df}{dr}\geq 0$. Posing $\frac{df}{dr}=f^{\prime }$, $%
\frac{dK}{dr}=K^{\prime }$, $\frac{dl}{dr}=l^{\prime }$ one has 
\begin{equation}
\omega _{1,2}=\frac{c}{l^{\prime }}\left( -K^{\prime }\pm \sqrt{K^{\prime
2}+l^{\prime }f^{\prime }}\right)  \label{geode}
\end{equation}

Introducing (\ref{geode}) into (\ref{cilindrico}) leads to

\begin{eqnarray*}
\Delta \tau &=&\frac{\pi l^{\prime }}{c\sqrt{K^{\prime 2}+l^{\prime
}f^{\prime }}}(\sqrt{f-2\frac{K}{l^{\prime }}\left( -K^{\prime }+\sqrt{%
K^{\prime 2}+l^{\prime }f^{\prime }}\right) -\frac{1}{l^{\prime }}\left(
-K^{\prime }+\sqrt{K^{\prime 2}+l^{\prime }f^{\prime }}\right) ^{2}}+ \\
&&-\sqrt{f-2\frac{K}{l^{\prime }}\left( -K^{\prime }-\sqrt{K^{\prime
2}+l^{\prime }f^{\prime }}\right) -\frac{1}{l^{\prime }}\left( -K^{\prime }-%
\sqrt{K^{\prime 2}+l^{\prime }f^{\prime }}\right) ^{2}})
\end{eqnarray*}

The condition for our two freely falling clocks to stay synchronous is (\ref
{rotante}), which, together with (\ref{geode}) gives

\begin{equation}
\frac{K^{\prime }}{l^{\prime }}=\frac{K}{l}  \label{kappa}
\end{equation}

This equation is identically satisfied whenever it is $K$\ proportional to $%
l $; otherwise it gives the special values of $r$, if they exist,\emph{\ }
such that the synchrony condition is maintained.

The result is again independent from any peculiar reference frame.

\section{Special cases}

The treatment of the problem has been quite general up to this moment. We
can of course specialize to some particular cases. Bonnor and Steadman
considered both the van Stockum space time\cite{kramer} and the Kerr metric.
They however focused on a quantity depending on the reference frame because
they compared different revolution times seen by an observer, rather than
proper time lapses between absolute events such as the coincidence of two
clocks in space and time.

The calculation of the absolute synchrony defect for a constant radius and
constant angular speed geodesic orbit around a rigidly rotating dust
cylinder (van Stockum metric) is straightforward but rather tedious and not
particularly useful. We may just explicit the condition for maintaining
synchrony, recovering from\cite{bonnor1} the following expressions for the
exterior of the cylinder (whose radius is R) and adapting them to our
notations: 
\begin{eqnarray*}
K &=&\frac{a}{c}\frac{R^{2}}{4n}\left[ \left( 2n+1\right) \left( \frac{r}{R}%
\right) ^{2n+1}+\left( 2n-1\right) \left( \frac{r}{R}\right) ^{1-2n}\right]
\\
l &=&\frac{R^{2}}{16n}\left[ \left( 2n+1\right) ^{3}\left( \frac{r}{R}%
\right) ^{2n+1}+\left( 2n-1\right) ^{3}\left( \frac{r}{R}\right)
^{1-2n}\right] \\
n^{2} &=&\frac{1}{4}-\frac{a^{2}R^{2}}{c^{2}}
\end{eqnarray*}

The parameter $a$ can be interpreted as the angular velocity on the axis of
the cylinder.

It is 
\begin{eqnarray*}
K^{\prime } &=&\frac{a}{c}\frac{R^{2}}{4n}\left[ \frac{\left( 2n+1\right)
^{2}}{r}\left( \frac{r}{R}\right) ^{2n+1}-\allowbreak \frac{\left(
2n-1\right) ^{2}}{r}\left( \frac{r}{R}\right) ^{1-2n}\right] \\
l^{\prime } &=&\frac{1}{16}\frac{R^{2}}{n}\left( \frac{\left( 2n+1\right)
^{4}}{r}\left( \frac{r}{R}\right) ^{2n+1}-\frac{\left( 2n-1\right) ^{4}}{r}%
\left( \frac{r}{R}\right) ^{1-2n}\right)
\end{eqnarray*}

and (\ref{kappa}) reduces to

\[
\frac{\left( 2n+1\right) ^{2}\left( \frac{r}{R}\right) ^{2n+1}-\allowbreak
\left( 2n-1\right) ^{2}\left( \frac{r}{R}\right) ^{1-2n}}{\left( 2n+1\right)
^{4}\left( \frac{r}{R}\right) ^{2n+1}-\left( 2n-1\right) ^{4}\left( \frac{r}{%
R}\right) ^{1-2n}}=\frac{\left( 2n+1\right) \left( \frac{r}{R}\right)
^{2n+1}+\left( 2n-1\right) \left( \frac{r}{R}\right) ^{1-2n}}{\left(
2n+1\right) ^{3}\left( \frac{r}{R}\right) ^{2n+1}+\left( 2n-1\right)
^{3}\left( \frac{r}{R}\right) ^{1-2n}} 
\]
or ($X=\left( \frac{r}{R}\right) ^{4n}$) 
\[
\frac{\left( 2n+1\right) ^{2}X-\allowbreak \left( 2n-1\right) ^{2}}{\left(
2n+1\right) ^{4}X-\left( 2n-1\right) ^{4}}=\frac{\left( 2n+1\right) X+\left(
2n-1\right) }{\left( 2n+1\right) ^{3}X+\left( 2n-1\right) ^{3}} 
\]

The formal solutions are 
\begin{eqnarray*}
r &=&R\left( \frac{2n-1}{2n+1}\right) ^{1/4n} \\
r &=&R\left( \frac{1-2n}{2n+1}\right) ^{3/4n}
\end{eqnarray*}
Provided $n$ is real, the upper solution holds when $\left| n\right| >\frac{1%
}{2}$, the other one when $\left| n\right| <\frac{1}{2}$. It is however $r<R$
in any case, this means that no exterior circular geodesic orbit exists
along which two clocks can stay synchronous.

Let us come now to the Kerr metric. A problem similar to the present one,
where instead of the two clocks we found a couple of light beams, has been
treated in \cite{tartaglia}. Using the same notations as there and
considering an equatorial circular orbit, the elementary arc length of the
fourdimensional helix of a clock is 
\[
d\tau =\sqrt{\frac{1}{\omega ^{2}}-\frac{a^{2}}{c^{4}}-\frac{r^{2}}{c^{2}}-2G%
\frac{M}{c^{2}r}\left( \frac{1}{\omega }-\frac{a}{c^{2}}\right) ^{2}}d\phi 
\]
The parameter $a=J/M$ is the ratio between the angular momentum $J$ and the
mass $M$ of the source.

Using the general formulae (\ref{span}) and (\ref{spannuno}) we obtain for
the proper times marked by the two clocks between two successive rendez
vous: 
\begin{eqnarray*}
\tau _{1} &=&\sqrt{1-\left( \frac{a^{2}}{c^{2}}+r^{2}\right) \frac{\omega
_{1}^{2}}{c^{2}}-2G\frac{M}{c^{2}r}\left( 1-\frac{a}{c^{2}}\omega
_{1}\right) ^{2}}\frac{2\pi }{\left| \omega _{2}-\omega _{1}\right| } \\
\tau _{2} &=&\sqrt{1-\left( \frac{a^{2}}{c^{2}}+r^{2}\right) \frac{\omega
_{2}^{2}}{c^{2}}-2G\frac{M}{c^{2}r}\left( 1-\frac{a}{c^{2}}\omega
_{2}\right) ^{2}}\frac{2\pi }{\left| \omega _{2}-\omega _{1}\right| }
\end{eqnarray*}
which amounts to have an absolute synchrony defect

\begin{eqnarray}
\Delta \tau &=&\frac{2\pi }{\left| \omega _{2}-\omega _{1}\right| }|\sqrt{%
1-\left( \frac{a^{2}}{c^{2}}+r^{2}\right) \frac{\omega _{2}^{2}}{c^{2}}-2G%
\frac{M}{c^{2}r}\left( 1-\frac{a}{c^{2}}\omega _{2}\right) ^{2}}+
\label{kerro} \\
&&-\sqrt{1-\left( \frac{a^{2}}{c^{2}}+r^{2}\right) \frac{\omega _{1}^{2}}{%
c^{2}}-2G\frac{M}{c^{2}r}\left( 1-\frac{a}{c^{2}}\omega _{1}\right) ^{2}}| 
\nonumber
\end{eqnarray}

The $\omega $'s are the ones seen by a far away inertial observer. To
maintain synchrony ($\Delta \tau =0$) it must be 
\begin{equation}
\omega _{2}=-\omega _{1}+\frac{4GMac^{2}}{a^{2}c^{2}r+r^{3}c^{4}+2GMa^{2}}
\label{nonrota}
\end{equation}

The last term on the right is twice the angular velocity of a ''locally non
rotating observer'' (LNRO)\cite{misner}. Actually such an observer would
indeed find that two counter-rotating clocks remain synchronous in his
reference frame: converting (\ref{nonrota}) into the variables of the LNRO
one obtains $\omega _{2}^{\prime }=-\omega _{1}^{\prime }$.

It is possible to specialize the result to circular geodesics in the
equatorial plane. It is then enough to substitute into (\ref{kerro}) the
angular speeds\cite{bonnor}: 
\begin{equation}
\omega _{1,2}=\frac{c^{2}}{a\pm c^{2}\sqrt{\frac{r^{3}}{GM}}}
\label{geocirco}
\end{equation}
The explicit results are not particularly enlightening though.

The solution of the problem in the Kerr metric lends itself to the study of
its weak field limit suitable for clocks moving around ordinary astronomical
spinning bodies such as the earth or the sun.

Let us assume for simplicity that 
\[
\frac{a}{cr}\sim \frac{\omega r}{c}\sim G\frac{M}{c^{2}r}<<1
\]
From (\ref{kerro}) we have consequently 
\begin{equation}
\Delta \tau \simeq 4\pi \frac{GM}{c^{4}r}a-\frac{\pi r^{2}}{c^{2}}\left(
\omega _{1}+\omega _{2}\right)   \label{questo}
\end{equation}

The second term is nothing but the first order approximation of (\ref{lack}%
). For a circular geodesic (\ref{geocirco}) gives approximately 
\[
\omega _{1}+\omega _{2}\simeq -2\frac{GM}{c^{2}r^{3}}a
\]
The final expression for (\ref{questo}) is then 
\begin{equation}
\Delta \tau \simeq 6\pi \frac{GM}{c^{4}r}a=6\pi \frac{GI}{c^{4}r}\Omega 
\label{finale}
\end{equation}

This result coincides with the one obtained in ref. \cite{lichten} provided
one remembers that here we refer to the first conjunction of the clocks
after the origin (a half revolution) whereas there the second one is
considered (a full revolution).

In the last term of (\ref{finale}) the non relativistic form for $a$
appears, being $I$ the classical moment of inertia of the source and $\Omega 
$ its angular speed (if a solid object is considered). 

\section{Solar system estimates}

Considering for simplicity solid, homogeneous, spherical, non relativistic
objects, it is\cite{tartaglia}: 
\begin{equation}
a=\frac{2}{5}R^{2}\Omega  \label{approssi}
\end{equation}
where $R$ is the radius of the body and $\Omega $ its angular velocity.

The numerical values of (\ref{approssi}) respectively for the Earth, the Sun
and Jupiter are 
\begin{eqnarray*}
a_{Earth} &=&1.2\times 10^{9}\text{ m}^{2}\text{/s} \\
a_{Sun} &=&8.9\times 10^{11}\text{m}^{2}\text{/s} \\
a_{Jupiter} &=&3.6\times 10^{11}\text{m}^{2}\text{/s}
\end{eqnarray*}
Correspondingly the orders of magnitude of the synchrony defects at the
first conjunction (after a half revolution) will be 
\begin{eqnarray*}
\Delta \tau _{Earth} &\sim &10^{-16}\text{ s} \\
\Delta \tau _{Sun} &\sim &10^{-11}\text{ s} \\
\Delta \tau _{Jupiter} &\sim &10^{-12}\text{ s}
\end{eqnarray*}

The radii of the orbits of the clocks have been assumed to be respectively $%
\sim 10^{7}$ m, $\sim 10^{10}$ m, $\sim 10^{8}$ m.

Time differences like these are extremely small but, at least for the Sun
and Jupiter they are in the range of measurability and their detection would
be a test of the influence that the angular momentum exerts on the pace of
the clocks. It must however be considered that the clocks to be compared
should be on board of orbiting spacecrafts and should be appropriately
stable for times sufficiently long.

A good parameter to measure the need for stability is obtained comparing the
synchrony defect at the first conjunction with the duration of one orbit.
For a typical satellite orbiting the Earth, this ratio is $\varepsilon \sim
10^{-19}$. 

In the case of Sun orbits of the order of the one of Mercury (period of a
couple months) correspond to $\varepsilon \sim 10^{-17}$.

The situation would be better for an artificial satellite of Jupiter (10
hours or so period) where $\varepsilon \sim 10^{-16}$.

\section{Conclusion}

We have shown that the most convenient definition of a gravitomagnetic
effect on clocks is based on two fiducial absolute events which are two
successive crossings of the four dimensional orbits of the clocks, when they
rotate about an axis. Basically the effect is topological in origin and it
is appropriate to refer to gravity only when the source of the field
possesses an angular momentum. Actually the Sagnac effect and its general
relativistic counterpart\cite{tartaglia} are but a special case of the
phenomenon we have described, when the two ''clocks'' are light beams.

Comparison of the readings of two clocks oppositely orbiting along circular
equatorial geodesics would be a means to reveal the dragging effect of the
spinning central body and to measure its angular momentum.

The numerical estimates we have made show that this effect within the solar
system is extremely weak, however it is also cumulative. Considering both
the size of the effect and the stability requirements the best condition
would be attained for a couple of clocks orbiting around Jupiter, also
because of the environment less hostile than around the Sun.

\end{document}